
\magnification=\magstep1
\overfullrule=0pt
\def\q#1{\lbrack #1 \rbrack}
\def\pano{\par\noindent}
\def\smno{\smallskip\noindent}
\def\meno{\medskip\noindent}
\def\bigno{\bigskip\noindent}
\def\o#1{\overline{#1}}
\def\pt{\partial}
\def\ts{\textstyle}
\def\cl{\centerline}
\def\id{1\hskip-2.5pt{\rm l}}
\def\section#1{\leftline{\bf #1}\vskip-7pt\line{\hrulefill}}
\def\bibitem#1{\parindent=8mm\item{\hbox to 6 mm{$\q{#1}$\hfill}}}
\def\BNT{\,\hbox{\hbox to -0.2pt{\vrule height 6.5pt width .2pt\hss}\rm N}}
\def\BRT{\,\hbox{\hbox to -0.2pt{\vrule height 6.5pt width .2pt\hss}\rm R}}
\def\BZT{{\rm Z{\hbox to 3pt{\hss\rm Z}}}}
\def\BZS{{\hbox{\sevenrm Z{\hbox to 2.3pt{\hss\sevenrm Z}}}}}
\def\BZSS{{\hbox{\fiverm Z{\hbox to 1.8pt{\hss\fiverm Z}}}}}
\def\BZ{{\mathchoice{\BZT}{\BZT}{\BZS}{\BZSS}}}
\def\BQT{\,\hbox{\hbox to -2.8pt{\vrule height 6.5pt width .2pt\hss}\rm Q}}
\def\BQS{\,\hbox{\hbox to -2.1pt{\vrule height 4.5pt width .2pt\hss}$
  \scriptstyle\rm Q$}}
\def\BQSS{\,\hbox{\hbox to -1.8pt{\vrule height 3pt width
    .2pt\hss}$\scriptscriptstyle \rm Q$}}

\def\BCT{\,\hbox{\hbox to -3pt{\vrule height 6.5pt width .2pt\hss}\rm C}}
\def\BCS{\,\hbox{\hbox to -2.2pt{\vrule height 4.5pt width .2pt\hss}$
    \scriptstyle\rm C$}}
\def\BCSS{\,\hbox{\hbox to -2pt{\vrule height 3.3pt width
    .2pt\hss}$\scriptscriptstyle \rm C$}}
\def\BC{{\mathchoice{\BCT}{\BCT}{\BCS}{\BCSS}}}
\def\figindents{\leftskip=3 true pc \rightskip=2 true pc}
\def\Verline#1#2#3{\rlap{\kern#1mm\raise#2mm
                   \hbox{\vrule height #3mm width 0.7pt depth 0 pt}}}
\def\Horline#1#2#3{\rlap{\kern#1mm\raise#2mm
                   \vbox{\hrule height 0.7pt width #3mm depth 0 pt}}}
\def\putbox#1#2#3{\setbox117=\hbox{#3}
                  \dimen121=#1mm
                  \dimen122=#2mm
                  \dimen123=\wd117
                  \dimen124=\ht117
                  \divide\dimen123 by -2
                  \divide\dimen124 by -2
                  \advance\dimen121 by \dimen123
                  \advance\dimen122 by \dimen124
                  \rlap{\kern\dimen121\raise\dimen122\hbox{#3}}}
\def\free{1}
\def\banks{2}
\def\self{3}
\def\selfb{4}
\def\cande{5}
\def\dine{6}
\def\disgre{7}
\def\diskacha{8}
\def\diskachb{9}
\def\egu{10}
\def\gepe{11}
\def\kava{12}
\def\oda{13}
\def\schwar{14}
\def\sche{15}
\def\schz{16}
\def\schd{17}
\def\silver{18}
\def\witten{19}
\def\wit{20}

\font\Large=cmr12 scaled \magstep3
\rm
\nopagenumbers
\pano
{\rightline {\vbox{\hbox{hep-th/9506104}
                   \hbox{BONN-TH-95-11}
                   \hbox{IFP-507-UNC}
                   \hbox{June 1995}}}}
\bigno\bigno
\centerline{\Large Exactly Solvable (0,2) Supersymmetric}
\vskip 10pt
\centerline{\Large String Vacua With GUT Gauge Groups }
\vskip 1.0cm
\centerline{{Ralph Blumenhagen${}^1$}\ \ and \ \
            {Andreas Wi{\ss}kirchen${}^2$}}
\vskip 1.0cm
\centerline{${}^1$ \it Institute of Field Physics, Department of Physics
and Astronomy,}
\centerline{\it University of North Carolina,
Chapel Hill NC 27599-3255, USA}
\vskip 0.1cm
\centerline{${}^2$ \it Physikalisches Institut der Universit\"at Bonn,
Nu{\ss}allee 12, 53115 Bonn, Germany}
\vskip 1.0cm
\centerline{\bf Abstract}
\meno
We present a construction of modular invariant partition functions for
heterotic $(0,2)$ supersymmetric classical string vacua. This
generalization of Gepner's construction yields GUT gauge groups $E_6$,
$SO(10)$, $SU(5)$ and $SU(3)\times SU(2)\times U(1)^r$, respectively.
By calculating the massless spectrum of some of these models we find
strong indications that they correspond to $(0,2)$ string vacua
discussed recently in the context of CYM/LG phases.
\footnote{}
{\pano
${}^1$ e-mail: blumenha@physics.unc.edu
\pano
${}^2$ e-mail: wisskirc@avzw01.physik.uni-bonn.de}
\vfill
\eject
\footline{\hss\tenrm\folio\hss}
\pageno=1
\section{1.\ Introduction}
\meno
Due to the lack of a nonperturbative formulation of string theory we are
still restricted to a perturbative search for reasonable string vacua. In
the last years classical solutions with $N=1$ space-time supersymmetry have
been studied intensively. A necessary condition is that the non flat
space-time directions are compactified on a Calabi-Yau manifold (CYM)
[\cande] or that the internal conformal field theory (CFT) has $(0,2)$
world sheet supersymmetry [\banks], respectively. Besides early indications
that generic $(0,2)$ string models might be destabilized by world sheet
instantons [\dine], the symmetric $(2,2)$ models are much easier to handle,
so that most effort focused only on their investigation. The implied
restriction for the nonlinear $\sigma$ model is that the spin connection is
identified with the gauge connection breaking one of the $E_8$ factors down
to $E_6$. However, ever since the work of Witten [\witten] in 1986 as well
as Distler and Greene [\disgre] in 1988, it has been known that weakening
the latter identification leads to more realistic GUT gauge groups like
$SO(10)$ or $SU(5)$. At the same time CFT models with different kinds
of gauge groups have been constructed using free bosons, free fermions or
orbifold techniques [\free]. In modern terminology those examples with
$N=1$ space-time supersymmetry are of type $(0,2)$. For the $(0,2)$ CY
models of Witten, Distler and Greene the left moving fermions of the
$\sigma$ model are not any longer sections of the tangent bundle of the
CYM, but of a more general stable holomorphic vector bundle of rank four
or five, respectively. Since there were those already mentioned reasonable
doubts about the consistency at all, these models lost their attraction
very fast.
\par
A revival of these models was initiated by Witten's work [\wit] on the
correspondence between nonlinear $\sigma$ models on CYMs and orbifolds of
Landau-Ginzburg (LG) models with isolated singularities. Applying his
techniques also to the $(0,2)$ case yielded a LG description which allowed
one to obtain more detailed information about the properties of a possible
conformal fixed point [\diskacha]. In [\diskachb] it was shown that at
least for marginal deformations of $(2,2)$ models by gauge singlets one
gets a bona fide CFT. Recently, Silverstein and Witten [\silver] argued
that even for all $(0,2)$ models described by linear $\sigma$ models, the
CFTs exist. Nevertheless, the explicit construction of such CFTs was still
unclear. Fortunately, some aspects of the structure of these CFTs can
already be explored in the LG framework [\diskacha]. Due to the left
moving R-invariance there exists a left moving $U(1)$ current, the
spectral flow operator of which extends $SO(8)$ to $SO(10)$ or $SO(6)$ to
$SU(5)$, respectively. Furthermore, the central charges for the left and
right moving sector can be calculated correctly; for $SO(10)$ they are
$(c,\o{c})=(10,9)$ and for $SU(5)$ one obtains $(c,\o{c})=(11,9)$. Thus,
it seems to be quite a tough problem to systematically construct modular
invariant partition functions for this class of purely heterotic CFTs.
\pano
In this paper we present a class of CFTs which satisfies all the conditions
mentioned above and which exhibits net numbers of generations which can be
reproduced using the CYM/LG framework. One well known way of building
new modular invariant partition functions is the simple current technique
developed by Schellekens and Yankielowicz [\sche,\schz]. In the spirit of
an idea mentioned in [\schd] we show that it is also suitable for the
construction of the desired $(0,2)$ string vacua. Using light cone gauge,
for $SO(10)$ we start with a diagonal invariant partition function of a
$(c,\o{c})=(24,24)$ CFT, which contains the four-dimensional space-time
part, an internal $(c,\o{c})=(9,9)$ $N=2$ supersymmetric part written as a
non supersymmetric CFT, a $U(1)_2$ part and the Kac-Moody algebra
${SO}(8)\times E_8$ of level one. Then we use simple current projections on
the right moving side to extend firstly $SO(8)\times U(1)_2$ to $SO(10)$.
This allows to apply the bosonic string map to yield a right moving
superstring with $\o{c}=12$. The idea of starting only with a subalgebra
of ${SO}(10)\times E_8$ and extending it has already been mentioned in
[\schd] but not carried out further. The new feature in our construction is
essentially the $U(1)_2$ factor. Afterwards, we project onto $NS$-$NS$ and
$R$-$R$ couplings guaranteeing that we choose the `supersymmetric tensor
product' on the r.h.s. The last operation to be carried out on the right
is the GSO projection onto even overall $U(1)$ charges. If we would stop at
this stage we would get nothing else but the usual Gepner models with $E_6$
gauge group [\gepe]. However, because of the new $U(1)_2$ factor there
occur new possibilities of preventing all the right moving operations to
act also on the left. Thus, we divide out the most complicated simple
current one can think of containing both pieces of $NS$ sectors and $R$
sectors. In general, this breaks the left moving $N=2$ supersymmetry and
the $E_6$ gauge group. The last step is to perform the left moving GSO
projection extending $SO(8)\times U(1)_{c=9}\times U(1)_2$ to
$SO(10)\times U(1)$. Note that this extension is different from the one
carried out on the r.h.s. Finally, we arrive at a modular invariant
partition function with gauge group $SO(10)$ and a $(c,\o{c})=(10,9)$ CFT
in the internal sector. Since there occur new combinations of left moving
excitations which are massless, in general the spectrum of the string
changes drastically. Comparing these spectra to those of the Distler/Kachru
models we found indications that we have really constructed CFTs,
describing certain points in the moduli space of the latter models.
\pano
The $SU(5)$ case is quite analogous, instead of one $U(1)_2$ CFT one uses
two such factors. As expected for consistent string vacua, the gauge
anomaly cancellation comes out automatically [\schwar]. Of course, the
above series can be extended further, three $U(1)_2$ factors generically
yield the non GUT gauge group $SU(3)\times SU(2)$. In general, in four
dimensions our construction allows all $E_r$ gauge groups with
$3\leq r\leq6$, which are defined by removing successively one simple
root of a long leg of the $E_6$ Dynkin diagram. In six and eight
space-time dimensions the usual Gepner construction yields the remaining
exceptional gauge groups $E_7$ and $E_8$, respectively.
\pano
This paper is organized as follows. In section 2 we review some basic
facts about the simple current technique. Then we present our construction
of $(0,2)$ modular invariant partition functions. The $SO(10)$ case is
discussed in detail, whereas $SU(5)$ and $SU(3)\times SU(2)$ are dealt
with rather briefly. A discussion of the general massless spectra follows
in section 6. In section 7 we present the results of a computer calculation
for some exemplary models like the quintic and compare them to the results
gained by CYM/LG techniques.
\bigno
\section{2.\ Review of the simple current technique}
\meno
This chapter contains only a very short review of the work of Schellekens
and Yankielowicz about generating new modular invariant partitions using
simple currents. For a more detailed discussion we refer the reader to the
original literature [\sche,\schz,\schd]. Suppose there is given a rational
conformal field theory (RCFT) with at least one modular invariant partition
function, e.g.\ the diagonal one. If this RCFT contains a simple current
$J$, i.e.\ $J\times\Phi_i=\Phi_j$ for every highest weight representation
 $\Phi_{i}$ of the chiral algebra in
the model, then one can obtain a new modular invariant in the following
way:\ First, define the index $N$ of the simple current $J$ to be the
smallest integer so that $J^N=\id$. Furthermore, the monodromy parameter
$r$ is determined by the conformal dimension of $J$:
$$ h(J)={r(N-1)\over 2N} \quad{\rm mod}\,1,\eqno(2.1) $$
so that $r$ is defined modulo $N$ for $N$ odd and modulo $2N$ for $N$ even.
Next, one defines the (monodromy) charge of a primary field $\Phi$:
$$ Q(\Phi)=h(\Phi)+h(J)-h(J\times \Phi)\quad{\rm mod}\,1,\eqno(2.2) $$
which takes values ${t\over N},\,t\in\BZ$. By the action of the simple
current all primaries of the RCFT are arranged in orbits
$\Phi,J\times\Phi,\ldots,J^d\times\Phi$, where $d$ is a divisor of $N$.
The charges of the fields occurring in an orbit are
${t+rn\over N}\ {\rm mod}\,1$.
\pano
If one can choose $r$ to be even, one can form a new modular invariant
partition function
$$Z(\tau,\o\tau)=\sum_{k,l}\chi_k(\tau)\,M_{kl}\,\chi_l(\o\tau)\eqno(2.3)$$
with the matrix $M$ determined by the orbits and the charges of the fields
with respect to the simple current $J$:
$$ M_{kl}=\sum_{p=1}^N \delta(\Phi_k,J^p \Phi_l)\,
     \delta^1\left(\hat Q(\Phi_k)+\hat Q(\Phi_l) \right),\eqno(2.4)$$
where $\delta^1(x)=1$ for $x\in\BZ$ and zero otherwise.
The slightly modified charge $\hat Q$ is defined on each orbit by
$$ \hat Q(J^n\Phi)={t+rn \over 2N} \quad{\rm mod}\,1.\eqno(2.5) $$
Two different kinds of invariants occur. On the one hand, those that result
from simple currents of integer conformal dimension. These can be regarded
as diagonal invariants for a subset of orbits having integer monodromy
charge. Thus, some of the original representations are really projected
out. On the other hand, simple currents of non integer dimension lead to
invariants corresponding to automorphisms of the fusion algebra which in
particular means that only the pairing of the left and right moving sector
changes.
\pano
Obviously, the product of two matrices (2.4) also defines a modular
invariant partition function which can be divided consistently by an
integer in order to guarantee the vacuum to appear only once. Thus, in
general one is allowed to form partition functions like
$$ Z(\tau,\o\tau)\sim \vec\chi(\tau)\, M(J_n)\ldots M(J_2)\, M(J_1)\,
              \vec{\chi}(\o\tau).\eqno(2.6) $$
The method of simple currents provides one with a powerful laboratory for
the construction of modular invariant partition functions. In [\schd] it
has extensively been used for the construction of four-dimensional, $N=1$
space-time supersymmetric string vacua with an internal $(c,\o{c})=(9,9)$
CFT. There, it already appeared that in general one gets only $(0,2)$ world
sheet supersymmetry. However, unlike to our models the gauge group
$SO(10)$ is linearly realized and not a result of a left moving GSO
projection implying the internal left moving central charge to be also
$c=9$ and not $c=10$. In the following sections we investigate whether this
large laboratory can also provide us with models of the Distler/Kachru
type [\diskacha].
\bigno
\section{3.\ String models with (0,2) supersymmetry and gauge group SO(10)}
\meno
In this section we make use of the simple current technique to find modular
invariant partition functions which satisfy all the properties known for
the conformal fixed points of the $(0,2)$ string vacua. First, we
concentrate on the $SO(10)$ case resulting from choosing a stable vector
bundle of rank four. In [\diskacha] the following information about the
CFT has been extracted from an LG analysis:
\smno
\item{$(a)$} The left and right conformal anomalies of the internal CFT are
             $(c,\o{c})=(10,9)$.
\smno
\item{$(b)$} Besides the right moving $U(1)$ current which is part of the
             right moving $N=2$ Virasoro algebra there exists a left
             moving $U(1)$ current satisfying the following operator
             product expansion (OPE):
             $$ J(z)\,J(w)={4\over (z-w)^2} +{\rm reg.} \eqno(3.1)$$
\smno
\item{$(c)$} Only the subset $SO(8)\times U(1)\subset SO(10)$ of the gauge
             group is linearly realized, the remaining roots are generated
             by taking orbits with respect to the spectral flow of
             conformal dimension $(h,q)=({1\over2},2)$.
\smno
Furthermore, we know that there is still a CYM in the model. As we have
learned from Gepner's work on $(2,2)$ models [\gepe], some of them
correspond to tensor products of unitary $N=2$ models. Using light cone
gauge we start with the diagonal partition function for the
$(c,\o{c})=(24,24)$ CFT model shown in Table 3.1.
\bigno
\cl{\vbox{
\hbox{\vbox{\offinterlineskip
\def\tablespace{height2pt&\omit&&\omit&&\omit&\cr}
\def\tablerule{\tablespace\noalign{\hrule}\tablespace}
\hrule\halign{&\vrule#&\strut\hskip0.2cm\hfil#\hfill\hskip0.2cm\cr
\tablespace
& part && $c$ && $\o{c}$ &\cr
\tablerule\tablerule
& $4D$ space-time, $X^{\mu}$ && $2$ && $2$ & \cr
\tablerule
& $N=2$ Virasoro && $9$ && $9$ &\cr
\tablerule
& $U(1)_2$ && $1$ && $1$ &\cr
\tablerule
& gauge group\ $SO(8)\times E_8$ && $12$ && $12$ &\cr
\tablespace}\hrule}}
\hbox{\hskip 0.5cm Table 3.1 \hskip 0.5cm Underlying CFT for $SO(10)$}}}
\meno
The remarkable change compared to Gepner's models is the appearance of a
free boson compactified on a circle of radius $R=2$ denoted as $U(1)_2$.
The diagonal partition function for this part can easily be expressed in
terms of $\Theta$-functions:
$$ Z_{U(1)_2}(\tau,\o\tau)=\sum_{m=-1}^2\Theta_{m,2}(\tau)\,
    \Theta_{m,2}(\o\tau).\eqno(3.2) $$
Note, that this is nothing else but the partition function of a Dirac
fermion. The fusion rules are quite simple:
$$[\Phi_{m,2}]\times[\Phi_{n,2}]=[\Phi_{m+n,2}]\quad{\rm mod}\,4.
  \eqno(3.3)$$
The current is $j_{U(1)_2}=i\pt\phi$ and satisfies the following OPE:
$$ j_{U(1)_2}(z)\,j_{U(1)_2}(w)={1\over (z-w)^2} +{\rm reg.} \eqno(3.4)$$
Furthermore, even though ${U(1)_2}$ is surely not $N=2$ supersymmetric,
there exists a spectral flow between the sector of even index $m$ and odd
index $m$ or between the $NS$ sector and the $R$ sector of the Dirac
fermion, respectively. The spectral flow operator is
${\rm exp}({i\phi(z)\over2})$ and has conformal dimension and charge
$(h,q)=({1\over8},{1\over2})$. Now, it becomes obvious why we have chosen
this special $c=1$ theory. Combining it with the left moving $c=9$ theory
offers the possibility to define an overall $U(1)$ current $J$ which
satisfies the conditions $(b)$ and $(c)$. The sum of the $N=2$ current
$j_{c=9}=i\sqrt{3}\pt\Phi(z)$ and the $U(1)_2$ current satisfies the OPE
in $(b)$ and the left moving spectral flow operator is given by
$$\Sigma_{c=10}(z)=e^{i{\sqrt{3}\over2}\Phi(z)}\otimes
     e^{i{1\over 2}\phi(z)}.\eqno(3.5)$$
Later on we will see that taking orbits with respect to this spectral flow
operator really extends $SO(8)\times U(1)$ to $SO(10)$.
\pano
Now we proceed by discussing the right moving sector. To this end let us
recall some facts about the representations of $SO(2n)$ Kac-Moody
algebras at level $k=1$. All we need for the following discussion is
summarized in Table 3.2.
\bigno
\cl{\vbox{
\hbox{\vbox{\offinterlineskip
\def\tablespace{height2pt&\omit&&\omit&&\omit&&\omit &\cr}
\def\tablerule{\tablespace\noalign{\hrule}\tablespace}
\hrule\halign{&\vrule#&\strut\hskip0.2cm\hfil#\hfill\hskip0.2cm\cr
\tablespace
& character && $h$ && $q$ mod 2 && degeneracy &\cr
\tablerule\tablerule
& $\chi_0={1\over 2}\left( \left({\vartheta_3\over\eta}\right)^n+
 \left({\vartheta_4\over\eta}\right)^n\right)$ && $0$ && $0$ && $0$ &\cr
\tablespace
& $\chi_v={1\over 2}\left( \left({\vartheta_3\over \eta}\right)^n-\left(
 {\vartheta_4\over\eta}\right)^n\right)$ && ${1\over2}$ && $1$ && $2n$ &\cr
\tablespace
& $\chi_s={1\over 2}\left( \left({\vartheta_2\over\eta}\right)^n+\left(
 {\vartheta_1\over\eta}\right)^n\right)$ && ${n\over8}$ && ${n\over2}$ &&
 $2^{n-1}$ & \cr
\tablespace
& $\chi_c={1\over 2} \left( \left({\vartheta_2\over\eta}\right)^n-\left(
  {\vartheta_1\over\eta}\right)^n\right)$ && ${n\over8}$ && ${n\over2}-1$
 && $2^{n-1}$& \cr
\tablespace}\hrule}}
\hbox{\hskip 0.5cm Table 3.2 \hskip 0.5cm Representations of $SO(2n)_1$}}}
\meno
The charge $q$ is taken with respect to the sum of all Cartan elements of
the Lie algebra $SO(2n)$ and $\vartheta_i$ denotes the Jacobi
$\vartheta$-functions. The fusion rules for the representations are
different for $n$ even and $n$ odd as one can read off from Table 3.3.
\bigno
\cl{\vbox{
\hbox{\vbox{\offinterlineskip
\def\tablespace{height2pt&\omit&&\omit&&\omit&&\omit &\cr}
\def\tablerule{\tablespace\noalign{\hrule}\tablespace}
\halign{\vrule\quad\hfill#\hfill&\strut\vrule#&\quad#\hfill\quad&\quad
      #\hfill\quad&\quad#\hfill\quad&\quad#\hfill\quad\vrule\cr
\noalign{\hrule}
$n$ odd && $0$ & $v$ & $s$ & $c$ \cr
\noalign{\hrule}
$0$ && $0$ & $v$ & $s$ & $c$ \cr
$v$ && $v$ & $0$ & $c$ & $s$ \cr
$s$ && $s$ & $c$ & $v$ & $0$ \cr
$c$ && $c$ & $s$ & $0$ & $v$ \cr
\noalign{\hrule}
}}
\vbox{\offinterlineskip
\def\tablespace{height2pt&\omit&&\omit&&\omit&&\omit &\cr}
\def\tablerule{\tablespace\noalign{\hrule}\tablespace}
\halign{\vrule\quad\hfill#\hfill&\strut\vrule#&\quad#\hfill\quad&\quad
      #\hfill\quad&\quad#\hfill\quad&\quad#\hfill\quad\vrule\cr
\noalign{\hrule}
$n$ even && $0$ & $v$ & $s$ & $c$ \cr
\noalign{\hrule}
$0$ && $0$ & $v$ & $s$ & $c$ \cr
$v$ && $v$ & $0$ & $c$ & $s$ \cr
$s$ && $s$ & $c$ & $0$ & $v$ \cr
$c$ && $c$ & $s$ & $v$ & $0$ \cr
\noalign{\hrule}
}}}
\hbox{\hskip 2.5cm Table 3.3 \hskip 0.5cm Fusion rules for $SO(2n)_1$}}}
\meno
In order to apply the bosonic string map for $SO(10)\times E_8\to SO(2)$
$$\eqalignno{\chi^{SO(10)\times E_8}_0 &\to\chi^{ SO(2)}_{v},\quad\quad
 \phantom{-}\chi^{ SO(10) \times E_8}_v \to\chi^{ SO(2) }_{0} &(3.6)\cr
 \chi^{SO(10) \times E_8}_s&\to -\chi^{ SO(2)}_{c}, \quad\quad
 \chi^{ SO(10) \times E_8}_{c}\to -\chi^{SO(2)}_{s} &\cr }$$
on the r.h.s.\ we have to extend $SO(8)\times U(1)_2$ to $SO(10)$. This can
be done by using the simple current
$$J_{(1\times8\to10)}=\Phi^{U(1)_2}_{2,2}\otimes
 \Phi^{SO(8)}_v,\eqno(3.7)$$
which generates the following orbits:
$$\eqalignno{
\chi^{SO(10)}_0&=\chi^{SO(8)}_0\Theta_{0,2}+\chi^{SO(8)}_v\Theta_{2,2}&\cr
\chi^{SO(10)}_v&=\chi^{SO(8)}_0\Theta_{2,2}+\chi^{SO(8)}_v\Theta_{0,2}&
    (3.8)\cr
\chi^{SO(10)}_s&=\chi^{SO(8)}_s\Theta_{1,2}+\chi^{SO(8)}_c\Theta_{-1,2}&\cr
\chi^{SO(10)}_c&=\chi^{SO(8)}_c\Theta_{1,2}+\chi^{SO(8)}_s\Theta_{-1,2}.&
\cr}$$
Now we can proceed on the right moving side in the same way as in Gepner's
construction. For the $c=9$ part we choose tensor products of unitary
representations of the $N=2$ super Virasoro algebra,
$$\eqalignno{ &c={3k\over k+2},\quad
              h^{l}_{m,s}={l(l+2)-m^2\over 4(k+2)}+{s^2\over 8},\quad
              q^{l}_{m,s}=-{m\over k+2}+{s\over 2},&(3.9)\cr
             & k\in\BNT,\ \ 0\leq l\leq k,\ \ -1\leq s\leq 2,\ \
             -l+\epsilon\leq m\leq l+\epsilon,\ \ l+m+s=0\ {\rm mod}\,2
&\cr }$$
with $\epsilon=0$ for $s\in\lbrace0,2\rbrace$ ($NS$ sector) and
$\epsilon=1$ for $s\in\lbrace-1,1\rbrace$ ($R$ sector), respectively.
Here we have split the characters in the usual way into two non
supersymmetric pieces:
$$ \chi^l_m=\chi^l_{m,s}+\chi^l_{m,s+2}.\eqno(3.10)$$
Then, in order to ensure that we are actually dealing with an $N=2$
supersymmetric model, we have to impose further projections
$$ J_i=G_i\otimes\Phi^{SO(8)}_v,\eqno(3.11)$$
where $G_i$ means the supercurrent in the $i$-th factor of the tensor
product. These projections allow only couplings between same kinds of
sectors. The last step to be carried out on the r.h.s.\ is the right
moving GSO projection onto states with even overall charge. The
necessary simple current is
$$J_{GSO_R}=\Sigma_{c=9}\otimes
 \Phi^{U(1)_2}_{1,2}\otimes\Phi^{SO(8)}_s.\eqno(3.12)$$
where $\Sigma_{c=9}$ denotes the spectral flow operator of
dimension $(h,q)=({3\over8},{3\over2})$ of the internal $c=9$ CFT.
In a concrete model $\Sigma_{c=9}$ simply contains one $\Phi^0_{1,1}$
primary field for each factor. So far, the partition function looks like
$$ Z\sim\vec{\chi}(\tau)\,M(J_{GSO_R})\,\prod_i M(J_i)\,
  M(J_{(1\times 8\to 10)})\,\vec{\chi}(\o\tau),\eqno(3.13)$$
which produces exactly the ordinary Gepner models, for all projections act
also on the left. Thus, in order to get something new we have to prevent
this by introducing more simple currents from the left which do not commute
with the simple currents in (3.13). Which simple currents are suitable
depends on the concrete model one is dealing with. However, on account of
the new $U(1)_2$ factor there occur simple currents which are not present
in the Gepner case. In general we are interested in simple currents which
both break the left moving $N=2$ supersymmetry and the $E_6$ gauge group
resulting from the $J_{GSO_R}$ projection. Suppose now we have found such
fields $\Upsilon_l$. What remains is only the left moving GSO
projection which is performed by the simple current
$$ J_{GSO_L}=\Sigma_{c=10}\otimes\Phi^{SO(8)}_s,\eqno(3.14)$$
which is actually the same as for the right moving GSO projection. However,
since the simple currents $J_i$ and $J_{(1\times 8\to 10)}$ do not any
longer act on the left, it does not yield an extension of the gauge group
to $E_6$ but only to $SO(10)$. On the level of characters this can be seen
by using a general theorem about orbits of spectral flows of chiral
dimension $(h,q)=({k\over2},k)$ [\oda]. Since in the $NS$ sector all
orbits contain only states with integral charge, Hermite's lemma
\footnote{${}^{\dag}$}
 {For $a\in\BNT$ and $0\leq b\leq a$ and $\delta=\pm 1$ be
 fixed:\ If $f(z)=f(z,q)$ is a Laurent series in $z$ and satisfies
 $f(zq,q)={\delta\over z^a q^{b\over2}}f(z,q)$, then $\lbrace f(z)\rbrace$
 is an $a$-dimensional vector space and one can choose the following
 basis:\ $z^{\rho}\sum_{n\in\BZ} \delta^n z^{an}
 q^{{a\over2}n^2+\left(\rho+{(b-a)\over 2}\right)n }$
 with $\rho=0,1,\ldots,a-1$.}
tells us that every orbit can be expanded into a finite number of $z$
dependent functions
$$f_{Q,k}(q,z)={1\over\eta(q)}\sum_{n\in\BZ}q^{{k\over 2}\left(
  n+{Q\over k}\right)^2}\,z^{{k}\left(n+{Q\over k}\right)},\quad Q\
  {\rm mod}\, k \eqno(3.15)$$
where the coefficients depend only on the variable $q$. In our case the
chiral spectral flow is twice the flow $\Sigma_{c=10}$ and therefore has
dimension $(h,q)=(2,4)$. Consequently, there are only four invariant
functions which can also be written in terms of $\Theta$ functions:
$$ f_{i,4}(q,z)={1\over\eta(q)}\left(\Theta_{2i,8}(q,z)+
  \Theta_{2(i+4),8}(q,z)\right),\quad -1\leq i\leq2.\eqno(3.16)$$
Note, that $f_{0,4},f_{2,4}$ have even charge and $f_{-1,4},f_{1,4}$ odd
charge. Since $\Sigma_{c=10}$ acts on the invariant functions by
$$ \Sigma_{c=10}:f_{i,4}\to f_{i+2,4},\eqno(3.17)$$
every orbit under $J_{GSO_L}$ with even charge can be expanded in the
following way:
$$\eqalignno{
 \chi^{j}_{orb}=&\chi_0^{SO(8)}\left[f_{0,4}A^j_0+f_{2,4}A^j_2\right]+
  \chi_v^{SO(8)}\left[ f_{1,4} A^j_1 + f_{-1,4} A^j_{-1}\right]&(3.18)\cr
  &\chi_s^{SO(8)}\left[ f_{2,4} A^j_0 + f_{0,4} A^j_2\right] +
  \chi_c^{SO(8)}\left[ f_{-1,4} A^j_1 + f_{1,4} A^j_{-1}\right].&\cr }$$
Reordering yields
$$\eqalignno{ \chi^{j}_{orb}=
 &\left[\chi_0^{SO(8)}f_{0,4}+\chi_s^{SO(8)}f_{2,4}\right] A^j_0 +
 \left[\chi_0^{SO(8)}f_{2,4}+\chi_s^{SO(8)}f_{0,4}\right] A^j_2+&(3.19)\cr
 &\left[\chi_v^{SO(8)}f_{1,4}+\chi_c^{SO(8)}f_{-1,4}\right] A^j_1 +
 \left[\chi_v^{SO(8)}f_{-1,4}+\chi_c^{SO(8)}f_{1,4}\right] A^j_{-1}.&\cr}$$
After some algebra neglecting the $z$ dependence this can be written as
$$\eqalignno{\chi^{j}_{orb}&=\chi_0^{SO(10)}A^j_0+\chi_v^{SO(10)}A^j_2+
  \chi_s^{SO(10)}A^j_1+\chi_c^{SO(10)}A^j_{-1} &(3.20)\cr}$$
showing explicitly the extension of the gauge group to $SO(10)$.
Summarizing, the entire model has the form
$$ Z\sim\vec{\chi}(\tau)\,M(J_{GSO_L})\,\prod_l M(\Upsilon_l)\,\,
  M(J_{GSO_R})\,\prod_i M(J_i)\,\,M(J_{(1\times 8\to 10)})\,
  \vec{\chi}(\o\tau)\eqno(3.21) $$
and by construction exhibits all the properties required at the beginning
of this section. The remaining question is whether one can really find
simple currents $\Upsilon_l$ which break both the left moving supersymmetry
and the $E_6$ gauge group. An explicit computer calculation shows that
generically this is not difficult. Apparently, at least for the moment we
have no other criteria to decide what the influence of a set of simple
currents $\Upsilon_l$ is than to perform the explicit calculation. Of
course, those which act trivially on the $U(1)_2$ should correspond to
ordinary orbifold constructions of the CYM, whereas others reflect the
choice of a different vector bundle for the left moving $\sigma$ model
fermions and thus reducing the rank of the gauge group. In section 7 we
present some first results of an explicit calculation showing what kinds
of spectra one can expect from the models in (3.21).
\bigno
\section{4.\ String models with (0,2) supersymmetry and gauge group SU(5)}
\meno
The generalization of the above construction to $SU(5)$ is straightforward,
so that it will be presented more briefly. The CYM/LG analysis
reveals the following information about the conformal fixed point:
\smno
\item{$(a)$} The left and right conformal anomalies of the internal CFT are
             $(c,\o{c})=(11,9)$.
\smno
\item{$(b)$} The OPE of the left moving $U(1)$ current is
             $$ J(z)\,J(w)={5\over (z-w)^2} +{\rm reg.} \eqno(4.1)$$
\smno
\item{$(c)$} The subset $SO(6)\times U(1)\subset SU(5)$ is linearly
             realized, orbits with respect to the spectral flow of
             conformal dimension $(h,q)=({5\over8},{5\over2})$ generate
             the missing roots.
\smno
Analogously to the former case, we suggest the ansatz for a relevant model
presented in Table 4.1.
\bigno
\cl{\vbox{
\hbox{\vbox{\offinterlineskip
\def\tablespace{height2pt&\omit&&\omit&&\omit&\cr}
\def\tablerule{\tablespace\noalign{\hrule}\tablespace}
\hrule\halign{&\vrule#&\strut\hskip0.2cm\hfil#\hfill\hskip0.2cm\cr
\tablespace
& part && $c$ && $\o{c}$ &\cr
\tablerule\tablerule
& $4D$ space-time, $X^{\mu}$ && $2$ && $2$ & \cr
\tablerule
& $N=2$ Virasoro && $9$ && $9$ &\cr
\tablerule
& $U(1)_2\otimes U(1)_2$ && $2$ && $2$ &\cr
\tablerule
& gauge group\ $SO(6)\times E_8$ && $11$ && $11$ &\cr
\tablespace}\hrule}}
\hbox{\hskip 0.5cm Table 4.1 \hskip 0.5cm Underlying CFT for $SU(5)$}}}
\meno
On the r.h.s the extension of $U(1)_2\times U(1)_2\times SO(6)$ to
$SO(10)$ can be achieved by the following two simple currents:
$$ \eqalignno{
 J^1_{1\times1\times6\to10}&=\Phi^{U(1)_2}_{2,2}
   \otimes\Phi^{U(1)_2}_{0,2}\otimes\Phi^{SO(6)}_v &(4.2)\cr
 J^2_{1\times1\times6\to 10}&=\Phi^{U(1)_2}_{0,2}
   \otimes\Phi^{U(1)_2}_{2,2}\otimes\Phi^{SO(6)}_v.&\cr}$$
The projections ensuring $N=2$ supersymmetry on the right moving side are
still given by
$$ J_i=G_i\otimes\Phi^{SO(6)}_v\eqno(4.3)$$
and the GSO projection leading to $N=1$ space-time supersymmetry is
$$ J_{GSO_R}=\Sigma_{c=9}\otimes
 \Phi^{U(1)_2}_{1,2}\otimes\Phi^{U(1)_2}_{1,2}
 \otimes\Phi^{SO(6)}_s.\eqno(4.4)$$
So far, the model looks like
$$ Z\sim\vec{\chi}(\tau)\,M(J_{GSO_R})\,\prod_i M(J_i)
   \, \prod_{j=1}^2 M(J^j_{(1\times 1\times 6\to 10)})\,
   \,\vec{\chi}(\o\tau).\eqno(4.5)$$
The left moving $U(1)$ current is the sum of the $N=2$ current and the two
$U(1)_2$ currents and in particular satisfies the OPE (4.1). The associated
spectral flow operator of dimension $(h,q)=({5\over8},{5\over2})$ is
$$\Sigma_{c=11}(z)=e^{i{\sqrt{3}\over2}\Phi(z)}\otimes
 e^{i{1\over2}\phi_1(z)}\otimes e^{i{1\over2}\phi_2(z)}.\eqno(4.6)$$
In order to perform the left moving GSO projection we use the simple
current
$$ J_{GSO_L}=\Sigma_{c=9}\otimes\Phi^{U(1)_2}_{1,2}
 \otimes\Phi^{U(1)_2}_{1,2}\otimes\Phi^{SO(6)}_s\eqno(4.7)$$
again. In the $NS$ sector there exist five series invariant with respect to
the square of the flow $\Sigma_{c=11}$
$$ f_{Q,5}(q,z)={1\over\eta(q)}\sum_{m\in\BZ}q^{{5\over 2}\left(m+{Q\over5}
  \right)^2}z^{5\left(m+{Q\over 5}\right)},\quad -2\leq Q\leq2.\eqno(4.8)$$
Similar to the Gepner case [\self,\egu] they are not of definite charge
parity, so that we have to use the decomposition into $\Theta$ functions
$$ f_{Q,5}(q,z)={1\over\eta(q)}\left(\Theta_{2Q,10}(q,z)+
 \Theta_{2Q+10,10}(q,z) \right)\quad{\rm mod }\,20.\eqno(4.9)$$
Taking into account the action of $\Sigma_{c=11}$ on the $\Theta$
functions:
$$ \Sigma_{c=11}:\Theta_{i,10}\to\Theta_{i+5,10}\quad{\rm mod}\,20,
   \eqno(4.10)$$
every orbit of even overall charge can be expanded in the following way:
$$\eqalignno{\chi^{j}_{orb}=&
 \chi_0^{SO(6)}\left[\Theta_{0,10}A^j_0+\Theta_{-8,10}A^j_1+\Theta_{8,10}
   A^j_{-1}+ \Theta_{4,10}A^j_2+\Theta_{-4,10}A^j_{-2} \right] +&\cr
 &\chi_v^{SO(6)}\left[\Theta_{10,10}A^j_0+\Theta_{2,10}A^j_1+\Theta_{-2,10}
   A^j_{-1}+\Theta_{-6,10}A^j_2+\Theta_{6,10}A^j_{-2}\right]+&\cr
 &\chi_s^{SO(6)}\left[\Theta_{5,10}A^j_0+\Theta_{-3,10}A^j_1+\Theta_{-7,10}
   A^j_{-1}+\Theta_{9,10}A^j_2+\Theta_{1,10}A^j_{-2}\right]+&(4.11)\cr
 &\chi_c^{SO(6)}\left[\Theta_{-5,10}A^j_0+\Theta_{7,10}A^j_1+\Theta_{3,10}
   A^j_{-1}+\Theta_{-1,10}A^j_2+\Theta_{-9,10}A^j_{-2}\right].&\cr}$$
Reordering and the fact that the characters of $SU(5)$ can be written as
sums over products of those of $SU(4)$ and $\Theta$ functions at level ten
yields
$$ \chi^{j}_{orb}= \chi^{SU(5)}_0 A^j_0 + \chi^{SU(5)}_{10} A^j_{+1} +
   \chi^{SU(5)}_{\o 5} A^j_{+2} + \chi^{SU(5)}_{\o{10}} A^j_{-1} +
   \chi^{SU(5)}_{5} A^j_{-2}.\eqno(4.12)$$
This shows very nicely the extension of the gauge group to $SU(5)$. Unlike
$E_6$ and $SO(10)$ there exist two different representations for chiral
space-time fermions, the $\bf{10}$ and the ${\bf\o{5}}$ which together
contains one generation of the standard model. As we will see in the
following sections the number of total generations in $\bf{10}$ and
${\bf\o{5}}$ in general are not the same, whereas the number of
net generations are. Thus, the whole partition function is
$$ Z\sim\vec{\chi}(\tau)\,M(J_{GSO_L})\,\prod_l M(\Upsilon_l)\,
   M(J_{GSO_R})\,\prod_i M(J_i)\,\prod_{j=1}^2
   M(J^j_{(1\times 1\times 6\to 10)})\,\vec{\chi}(\o\tau).\eqno(4.13)$$
Apparently, the whole construction can be extended further starting with
three copies of $U(1)_2$ and the group $SO(4)\times E_8$ which extends to
$SU(3)\times SU(2)\times E_8$. Since at our exactly solvable points there
are a lot $U(1)$ factors around, one might get the supersymmetric standard
model. However, analogously to the $(2,2)$ case these $U(1)$ factors are
believed to exist only at this special point of the moduli space of the CYM
and would be broken by a generic marginal deformation. Thus, very sensitive
fine tuning is necessary to choose such a special string vacuum.
\vfill
\eject
\bigno
\section{5.\ String models with (0,2) supersymmetry and gauge group
$SU(3)\times SU(2)$}
\meno
The next kind of models are those which exhibit gauge group
$E_3=SU(3)\times SU(2)$. At certain points of the moduli space
$E_3$ may be extended by at least some $U(1)$ factors,
so that the gauge group contains the standard model.
The ansatz is shown in Table 5.1.
\bigno
\cl{\vbox{
\hbox{\vbox{\offinterlineskip
\def\tablespace{height2pt&\omit&&\omit&&\omit&\cr}
\def\tablerule{\tablespace\noalign{\hrule}\tablespace}
\hrule\halign{&\vrule#&\strut\hskip0.2cm\hfil#\hfill\hskip0.2cm\cr
\tablespace
& part && $c$ && $\o{c}$ &\cr
\tablerule\tablerule
& $4D$ space-time, $X^{\mu}$ && $2$ && $2$ & \cr
\tablerule
& $N=2$ Virasoro && $9$ && $9$ &\cr
\tablerule
& $U(1)_2\otimes U(1)_2\otimes U(1)_2$ && $3$ && $3$ &\cr
\tablerule
& gauge group\ $SO(4)\times E_8$ && $10$ && $10$ &\cr
\tablespace}\hrule}}
\hbox{\hskip 0.5cm Table 5.1 \hskip 0.5cm Underlying CFT for
$SU(3)\times SU(2)$}}}
\meno
On the r.h.s the extension of
$U(1)_2\times U(1)_2\times U(1)_2\times SO(4)$ to $SO(10)$ can be achieved
by the following three simple currents:
$$\eqalignno{
 J^1_{1^3\times4\to10}&=\Phi^{U(1)_2}_{2,2}\otimes
  \Phi^{U(1)_2}_{0,2}\otimes\Phi^{U(1)_2}_{0,2}
  \otimes\Phi^{SO(4)}_v &\cr
 J^2_{1^3\times4\to10}&=\Phi^{U(1)_2}_{0,2}\otimes
  \Phi^{U(1)_2}_{2,2}\otimes\Phi^{U(1)_2}_{0,2}
  \otimes\Phi^{SO(4)}_v &(5.1)\cr
 J^3_{1^3\times4\to10}&=\Phi^{U(1)_2}_{0,2}\otimes
  \Phi^{U(1)_2}_{0,2}\otimes\Phi^{U(1)_2}_{2,2}
  \otimes\Phi^{SO(4)}_v.&\cr }$$
The GSO projection is
$$ J_{GSO_R}=\Sigma_{c=9}\otimes
 \Phi^{U(1)_2}_{1,2}\otimes\Phi^{U(1)_2}_{1,2}\otimes
 \Phi^{U(1)_2}_{1,2}\otimes\Phi^{SO(4)}_s.\eqno(5.2)$$
This simple current is also used to perform the left moving GSO projection.
In the $NS$ sector there exist six series invariant with respect to
the square of the flow $\Sigma_{c=12}$:
$$ f_{Q,6}(q,z)={1\over\eta(q)}\sum_{m\in\BZ}q^{{3}\left(m+{Q\over 6}
 \right)^2}z^{6\left(m+{Q\over6}\right)},\quad -2\leq Q\leq3.\eqno(5.3)$$
Thus, every orbit of the left moving GSO projection can be expanded in the
following way:
$$\eqalignno{\chi^{j}_{orb}=
&\chi_0^{SO(4)}\left[f_{0,6}A^j_0+f_{2,6}A^j_2+f_{-2,6}A^j_{-2}\right]+&\cr
&\chi_v^{SO(4)}\left[f_{3,6}A^j_3+f_{1,6}A^j_1+f_{-1,6}A^j_{-1} \right]+&
  (5.4)\cr
&\chi_s^{SO(4)}\left[f_{3,6}A^j_0+f_{-1,6}A^j_2+f_{1,6}A^j_{-2}\right]+&\cr
&\chi_c^{SO(4)}\left[f_{0,6}A^j_3+f_{-2,6}A^j_{1}+f_{2,6}A^j_{-1}\right]
&\cr}$$
which can be rewritten in terms of $E_3$ characters
$$ \chi^{j}_{orb}=\chi^{E_3}_{(0,0)}A^j_0+\chi^{E_3}_{(0,2)}A^j_{3}+
        \chi^{E_3}_{(3,0)}A^j_{-2}+\chi^{E_3}_{(\o3,0)}A^j_{2}+
        \chi^{E_3}_{(3,2)}A^j_{-1}+\chi^{E_3}_{(\o3,2)}A^j_{1}.\eqno(5.5)$$
Thus, there appear four chiral representations. The whole partition
function is
$$ Z\sim\vec{\chi}(\tau)\,M(J_{GSO_L})\,\prod_l M(\Upsilon_l)\,
 M(J_{GSO_R})\,\prod_i M(J_i)\,\prod_{j=1}^3 M(J^j_{1^3\times 4 \to10})
      \,\vec{\chi}(\o\tau).\eqno(5.6)$$
Remarkably, the above generalized Gepner type construction yields
all gauge groups $E_6$, $E_5=SO(10)$, $E_4=SU(5)$ and
$E_3=SU(3)\times SU(2)$. The extension of the gauge group $SO(2n)\times
U(1)$ to $E_{n+1}$ is schematically represented by the extension
of the Dynkin diagrams shown in Figure 5.1.
\bigno
$$
  \putbox{0}{0}{$\circ$}
  \Horline{0.5}{0}{3.9}
  \putbox{5}{0}{$\circ$}
  \Horline{5.5}{0}{3.5}
  \putbox{11.5}{0}{$\cdots$}
  \Horline{13.9}{0}{3.5}
  \putbox{18}{0}{$\circ$}
  \Horline{18.5}{0}{3.9}
  \putbox{23}{0}{$\circ$}
  \Verline{17.8}{0.6}{3.9}
  \putbox{18}{5.1}{$\circ$}
  \putbox{39}{0}{$\buildrel U(1)\,\&\,spec.\,fl.\over\longrightarrow$}
  \putbox{11.5}{-5}{$D_n$}
  \putbox{54}{0}{$\circ$}
  \Horline{54.5}{0}{3.9}
  \putbox{59}{0}{$\circ$}
  \Horline{59.5}{0}{3.5}
  \putbox{65.5}{0}{$\cdots$}
  \Horline{67.9}{0}{3.5}
  \putbox{72}{0}{$\circ$}
  \Horline{72.6}{0}{3.8}
  \putbox{77}{0}{$\circ$}
  \Horline{77.6}{0}{3.8}
  \putbox{82}{0}{$\circ$}
  \Verline{71.85}{0.6}{3.9}
  \putbox{72}{5.1}{$\circ$}
  \putbox{65.5}{-5}{$E_{n+1}$}
  \hskip 8.5cm \hfill
$$
\bigno
{\figindents\hskip3.5cm Figure 5.1\hskip0.5cm Extension of Dynkin diagrams}
\bigno
In order to extract more concrete information about these models we have to
calculate some physical quantities. As a first step we concentrate on the
massless spectrum, especially the number of generations which can also be
calculated in the CYM/LG scheme.
\bigno
\section{6.\ The massless spectrum}
\meno
To begin with, there are the universal massless particles of the heterotic
string like the graviton, the gravitino and the gluons and gluinos of the
gauge group. The coupling of the vacuum on the l.h.s.\ to the space-time
SUSY supercharges
$$ \left({\ts\,\o{h}={3\over8},\o{q}}\right)_{\,\o{c}=9}\otimes
\left(\Phi^{U(1)_2}_{1,2}\right)^{5-n}\otimes\Phi^{SO(2n)}_s\eqno(6.1) $$
on the r.h.s.\ determines the degree of supersymmetry. If there occur $k$
such states one actually deals with $N=k$ space-time supersymmetry. In
most examples discussed in section 7 we only have $N=1$ space-time
supersymmetry. However, a compactification on $K_3\times T^2$ yields also
$N=2$ supersymmetry which recently has received attention because of
duality relations to type II Calabi-Yau compactifications [\kava].
\pano
In addition, for $SO(10)$ the spectrum contains spin zero and
spin one particles in the singlet, vector and spinor representations and
their corresponding superpartners of spin one half. For the bosons their
quantum numbers in the internal $(c,\o c)=(10,9)$ CFT are listed in the
form $(h,q;\o{h},\o{q})$ in Table 6.1.
\bigno
\cl{\vbox{
\hbox{\vbox{\offinterlineskip
\def\tablespace{height2pt&\omit&&\omit&&\omit&&\omit&&\omit&\cr}
\def\tablerule{\tablespace\noalign{\hrule}\tablespace}
\hrule\halign{&\vrule#&\strut\hskip0.2cm\hfil#\hfill\hskip0.2cm\cr
\tablespace
& && ${\bf0}$ && ${\bf10}$ && ${\bf16}$ && ${\bf\o{16}}$ &\cr
\tablerule\tablerule
& spin $0$ && $(1,0;{1\over 2},\pm1)$ && $({1\over2},0;{1\over2},\pm1)$ &&
  $({1\over2},1;{1\over2},\pm1)$ && $({1\over2},-1;{1\over2},\pm1)$ & \cr
\tablerule
& spin $1$ && $(1,0;0,0)$ && $({1\over2},0;0,0)$ && $({1\over2},1;0,0)$ &&
  $({1\over 2},-1;0,0)$ & \cr
\tablespace}\hrule}}
\hbox{\hskip 0.5cm Table 6.1\hskip 0.5cm Massless spectrum for $SO(10)$}}}
\meno
Since the right moving sector is the same as in ordinary $(2,2)$ models,
the same combinations of primaries from the different factor models
contribute to the right moving part of the massless states.
Thus, in a concrete calculation one starts with these combinations on the
r.h.s.\ of the chain in (3.21) and follows their way through all the simple
currents to determine to which states they couple on the l.h.s. The
appearance of spin one particles in the spinor representations indicates
an extension of the gauge group, in the $SO(10)$ case usually to $E_6$.
However, in general this does not mean that also the $N=2$ supersymmetry
is restored in the left moving sector.
\pano
In the case of $SU(5)$ only the singlet and the four spinor representations
occur. The quantum numbers in the internal $(c,\o{c})=(11,9)$ CFT are
listed in Table 6.2.
\bigno
\cl{\vbox{
\hbox{\vbox{\offinterlineskip
\def\tablespace{height2pt&\omit&&\omit&&\omit&&\omit&&\omit&&\omit&\cr}
\def\tablerule{\tablespace\noalign{\hrule}\tablespace}
\hrule\halign{&\vrule#&\strut\hskip0.2cm\hfil#\hfill\hskip0.2cm\cr
\tablespace
& && ${\bf0}$ && ${\bf10}$ && ${\bf\o{10}}$ && ${\bf5}$ && ${\bf\o{5}}$&\cr
\tablerule\tablerule
& spin $0$ && $(1,0;{1\over2},\pm1)$
&& $({5\over8},-{3\over2};{1\over2},\pm1)$
&& $({5\over 8},{3\over 2};{1\over 2},\pm 1)$
&& $({5\over 8},{1\over 2};{1\over 2},\pm 1)$
&& $({5\over 8},-{1\over 2};{1\over 2},\pm 1)$ & \cr
\tablerule
& spin $1$ && $(1,0;0,0)$ && $({5\over 8},-{3\over 2};0,0)$
&& $({5\over 8},{3\over 2};0,0)$ && $({5\over 8},{1\over 2};0,0)$
&& $({5\over 8},-{1\over 2};0,0)$ & \cr
\tablespace}\hrule}}
\hbox{\hskip 0.5cm Table 6.2 \hskip 0.5cm Massless spectrum for $SU(5)$}}}
\meno
In this case further gluons can extend the gauge symmetry to $SO(10)$,
$E_6$ and also to $SU(6)$. Like in all string models where only Kac-Moody
algebras at level one occur, one has to use a mechanism for breaking the
GUT down to the standard model different from the spontaneous symmetry
breaking by attaching a nonzero vacuum expectation value to some Higgs
fields in the adjoint representation. For instance, if the fundamental
group of the CYM is nontrivial one has the possibility to use Wilson lines.
\pano
For the gauge group $E_3$ the massless spectrum has the
internal $(c,\o c)=(12,9)$ quantum numbers listed in Table 6.3.
\bigno
\cl{\vbox{
\hbox{\vbox{\offinterlineskip
\def\tablespace{height2pt&\omit&&\omit&&\omit&&\omit&&\omit&&\omit&&\omit
 &\cr}
\def\tablerule{\tablespace\noalign{\hrule}\tablespace}
\hrule\halign{&\vrule#&\strut\hskip0.07cm\hfil#\hfill\hskip0.02cm\cr
\tablespace
& && ${\bf0}=(0,0)$ && ${\bf2}=(2,0)$ && ${\bf3}=(3,0)$ &&
 ${\bf\o{3}}=(\o3,0)$ && ${\bf6}=(3,2)$ && ${\bf\o{6}}=(\o3,2)$
&\cr \tablerule\tablerule
& spin $0$ && $(1,0;{1\over2},\pm1)$
&& $({3\over4},0;{1\over2},\pm1)$
&& $({3\over 4},{1};{1\over 2},\pm 1)$
&& $({3\over 4},-{1};{1\over 2},\pm 1)$
&& $({3\over 4},-{2};{1\over 2},\pm 1)$
&& $({3\over 4},{2};{1\over 2},\pm 1)$ & \cr
\tablerule
& spin $1$ && $(1,0;0,0)$ && $({3\over 4},0;0,0)$
&& $({3\over 4},1;0,0)$ && $({3\over 4},-1;0,0)$
&& $({3\over 4},-2;0,0)$
&& $({3\over 4},2;0,0)$ &\cr
\tablespace}\hrule}}
\hbox{\hskip 0.5cm Table 6.3 \hskip 0.5cm Massless spectrum for
$SU(3)\times SU(2)$}}}
\meno
After this partially quite technical presentation of exactly solvable
$(0,2)$ CFTs, we calculate the massless spectra for some exemplary models
in the following section.
\bigno
\section{7.\ Examples}
\meno
Since the most frequently discussed example of a CYM is the quintic
hypersurface in $\BC{\rm P}^4$, we will also focus our attention on the
corresponding $N=2$ CFT of five copies of the $k=3$ unitary model,
denoted as $(3)^5$. As some first results of a computer calculation
we present how appropriate choices of simple currents yield $(0,2)$
models with gauge groups $E_6$, $SO(10)$, $SU(5)$, $SU(3)\times SU(2)$
and even $E_7$, $SU(7)$, $SU(6)$, $SO(12)$, $SU(6)\times SU(2)$ and
$SU(4)\times SU(2)$. Analogously to the exactly solvable $(2,2)$ string
vacua these generic gauge groups are extended further by a factor $G$
which is a product of groups of small rank, usually $U(1)$ factors. In
most cases these extensions are not written down explicitly in the
following.
\meno
$\bullet$ {\bf Gauge group $SO(10)$}
\smno
We start with a model like in Table 3.1 where we choose $(3)^5$ as the
internal $c=9$ part. Besides all the projections in (3.21) we only include
one further simple current (for simplicity we will not write down the
`$\otimes$' in the following)
$$ \Upsilon=\Phi^3_{0,-1}\otimes\left(\Phi^0_{0,0}\right)^4
 \otimes\Phi^{U(1)_2}_{1,2}\otimes\Phi^{SO(8)}_0\eqno(7.1) $$
of dimension $(h,q)=(1,0)$. Note, that (7.1) contains both factors from the
$NS$ and the $R$ sector. The resulting massless spectrum is given in Table
7.1.
\bigno
\cl{\vbox{
\hbox{\vbox{\offinterlineskip
\def\tablespace{height2pt&\omit&&\omit&&\omit&&\omit&&\omit&\cr}
\def\tablerule{\tablespace\noalign{\hrule}\tablespace}
\hrule\halign{&\vrule#&\strut\hskip0.2cm\hfil#\hfill\hskip0.2cm\cr
\tablespace
& && ${\bf0}$ && ${\bf10}$ && ${\bf16}$ && ${\bf\o{16}}$ &\cr
\tablerule\tablerule
& spin $0$ && $350$ && $74$ && $80$ && $0$ & \cr \tablerule
& spin $1$ && $7$ && $0$ && $0$ && $0$ & \cr
\tablespace}\hrule}}
\hbox{\hskip 0.5cm Table 7.1 \hskip 0.5cm $SO(10)$ model}}}
\meno
Since no further gluons appear, the gauge group is $SO(10)$ by
construction. Looking in more detail where the 80 generations are coming
from, one realizes that 60 of them are ordinary $N=2$ states from the
r.h.s.\ surviving all projections. However, the remaining 20 states arise
in some orbits of $\Upsilon$ and contain both a nontrivial contribution
of the $U(1)_2$ part and a mixing of $NS$ and $R$ states.
\pano
The numbers of particles in $\bf10$, $\bf16$ and $\bf\o{16}$ agree with a
model discussed in [\diskacha] which was defined on a complete intersection
in the weighted projective space $WP^5_{(1,1,1,1,2,2)}$.
Therefore our model could lie in the moduli space of that model which is
based on a totally different approach. That would imply that our model,
although built up on the $(3)^5$ tensor product, does not live on the
quintic CYM.
\pano
Like in ordinary Gepner models there exists an isomorphic CFT, for which
generations and antigenerations are interchanged. This mirror model can be
obtained from the original one simply by including the following simple
currents in front of the right moving states $\vec\chi(\o\tau)$ in (3.21):
$$\eqalignno{ &J^M_1=\Phi^3_{3,2} \Phi^0_{0,0}\ldots \Phi^0_{0,0}\,
  \Phi^{U(1)_2}_{0,2} \Phi^{SO(8)}_{0} &\cr
  &J^M_2=\Phi^0_{0,0} \Phi^3_{3,2}\ldots \Phi^0_{0,0}\, \Phi^{U(1)_2}_{0,2}
   \Phi^{SO(8)}_{0}  &\cr &\quad\quad \vdots &\cr
  &J^M_5=\Phi^0_{0,0} \Phi^0_{0,0}\ldots \Phi^3_{3,2}\, \Phi^{U(1)_2}_{0,2}
   \Phi^{SO(8)}_{0}  &(7.2)\cr
  &J^M_6=\Phi^0_{0,0} \Phi^0_{0,0}\ldots \Phi^0_{0,0}\, \Phi^{U(1)_2}_{2,2}
   \Phi^{SO(8)}_{0}  &\cr
  &J^M_7=\Phi^0_{0,0} \Phi^0_{0,0}\ldots \Phi^0_{0,0}\, \Phi^{U(1)_2}_{0,2}
   \Phi^{SO(8)}_{v}.  &\cr }$$
The generalization to the general case is straightforward. Thus, the class
of models investigated in this paper exhibits mirror symmetry. For the
class of linear $\sigma$ models in [\disgre] mirror symmetry has not been
established so far.
\pano
It is easy to calculate at least net numbers of generations for the
Distler/Kachru models. Due to [7,17] the defining data of stable,
holomorphic vector bundles $V=V_1, V_2$ are given by the exact sequence
$$ 0\to\ V\to \bigoplus_{a=1}^{r+M} {\cal O}(n_a)\to \bigoplus_{i=1}^{M}
            {\cal O}(m_i)\to 0. \eqno(7.3)$$
Here $r=3,4,5,6$ yields gauge group $E_{9-r}$, and the $n_a$ and $m_i$ are
positive integers chosen such that $c_1(V_1)=c_1(V_2)=0$ and the gauge
anomaly vanishes, i.e.\ $c_2(V_1)+c_2(V_2)=c_2(T)$ with $T$ denoting the
tangent bundle. The net numbers of generations are given by an index
theorem
$$  N_{gen}={1\over 2} | \int_{M} c_3(V_1) | \eqno(7.4)$$
with
$$c_3(V_1)=-{1\over3}\left(\sum_i m_i^3-\sum_a n_a^3\right)J^3.\eqno(7.5)$$
By stochastically calculating some of these numbers for the quintic
one realizes that only multiples of $5$ occur and $80$ really appears.
Of course, much more work has to be done to definitely identify this model
with a concrete $\sigma$ model. There are at least two candidates, for
the model could be based on the quintic CYM or on the weighted projective
space mentioned above.
\meno
$\bullet$ {\bf Gauge group $E_6$}
\smno
Even though the construction yields gauge group $SO(10)$ it may happen that
further extended gauge groups occur. For instance, without any further
simple current $\Upsilon$ one gets the ordinary $(2,2)$ Gepner model with
gauge group $E_6$. However, including the simple current
$$ \Upsilon=\Phi^3_{-3,0} \Phi^0_{1,1} \left( \Phi^0_{0,0}\right)^3\,
  \Phi^{U(1)_2}_{1,2}\,\Phi^{SO(8)}_s  \eqno(7.6)$$
destroys the left moving $N=2$ supersymmetry but the $E_6$ remains
unbroken. This can be read off explicitly from the $SO(10)$ massless
spectrum listed in Table 7.2.
\bigno
\cl{\vbox{
\hbox{\vbox{\offinterlineskip
\def\tablespace{height2pt&\omit&&\omit&&\omit&&\omit&&\omit&\cr}
\def\tablerule{\tablespace\noalign{\hrule}\tablespace}
\hrule\halign{&\vrule#&\strut\hskip0.2cm\hfil#\hfill\hskip0.2cm\cr
\tablespace
& && ${\bf0}$ && ${\bf10}$ && ${\bf16}$ && ${\bf\o{16}}$ &\cr
\tablerule\tablerule
& spin $0$ && $432$ && $102$ && $101$ && $1$ & \cr
\tablerule
& spin $1$ && $5$ && $0$ && $1$ && $1$ & \cr
\tablespace}\hrule}}
\hbox{\hskip 0.5cm Table 7.2 \hskip 0.5cm Gepner-like model}}}
\meno
Thus, this model lies in the enhanced $(0,2)$ moduli space of the quintic
and is presumably connected to the $(2,2)$ moduli space by marginal
deformations with $E_6$ singlets on the left moving side. Therefore, it is
very suggestive of the $SO(10)$ deformation of the quintic.
\pano
But one does not need to reproduce the Gepner spectrum. Dividing out the
simple current
$$ \Upsilon=\Phi^3_{-3,0} \Phi^0_{1,1} \left( \Phi^0_{0,0}\right)^3\,
  \Phi^{U(1)_2}_{1,2}\,\Phi^{SO(8)}_c  \eqno(7.7)$$
which is only slightly different from the current (7.6) one obtains a
totally different spectrum as given in Table 7.3.
\bigno
\cl{\vbox{
\hbox{\vbox{\offinterlineskip
\def\tablespace{height2pt&\omit&&\omit&&\omit&&\omit&&\omit&\cr}
\def\tablerule{\tablespace\noalign{\hrule}\tablespace}
\hrule\halign{&\vrule#&\strut\hskip0.2cm\hfil#\hfill\hskip0.2cm\cr
\tablespace
& && ${\bf0}$ && ${\bf10}$ && ${\bf16}$ && ${\bf\o{16}}$ &\cr
\tablerule\tablerule
& spin $0$ && $368$ && $62$ && $41$ && $21$ & \cr
\tablerule
& spin $1$ && $5$ && $0$ && $1$ && $1$ & \cr
\tablespace}\hrule}}
\hbox{\hskip 0.5cm Table 7.3 \hskip 0.5cm $E_6$ model}}}
\meno
$N_{gen}=20$ also appears in the list of Distler/Kachru spectra.
\meno
$\bullet$ {\bf Gauge group $SO(12)$}
\smno
It is known that the Gepner model $(1)^3(2)(6)^2$ describes a
compactification on $K_3\times T^2$ and yields both an extension of the
gauge group to $E_7$ and an enhanced $N=2$ space-time supersymmetry. In
[\kava] it has been argued that such models with at least $(0,4)$
world sheet supersymmetry have dual realizations as type II
compactifications on CYMs. Using the simple current
$$ \Upsilon= \Phi^1_{0,-1}\, \left( \Phi^0_{0,0}\right)^2\, \Phi^2_{0,0}
      \, \left( \Phi^0_{0,0}\right)^2\,
       \Phi^{U(1)_2}_{1,2} \,\Phi^{SO(8)}_0 \eqno(7.8)$$
in the $(1)^3(2)(6)^2$ model one obtains the spectrum in Table 7.4.
\bigno
\cl{\vbox{
\hbox{\vbox{\offinterlineskip
\def\tablespace{height2pt&\omit&&\omit&&\omit&&\omit&&\omit&\cr}
\def\tablerule{\tablespace\noalign{\hrule}\tablespace}
\hrule\halign{&\vrule#&\strut\hskip0.2cm\hfil#\hfill\hskip0.2cm\cr
\tablespace
& && ${\bf0}$ && ${\bf10}$ && ${\bf16}$ && ${\bf\o{16}}$ &\cr
\tablerule\tablerule
& spin $0$ && $220$ && $34$ && $12$ && $12$ & \cr
\tablerule
& spin $1$ && $14$ && $2$ && $0$ && $0$ & \cr
\tablespace}\hrule}}
\hbox{\hskip 0.5cm Table 7.4 \hskip 0.5cm $N=2$ and $SO(12)$ model}}}
\meno
This model has still $N=2$ space-time supersymmetry which requires $(0,4)$
world sheet supersymmetry. In addition, the two vectors in ${\bf10}$ extend
the gauge group to $SO(12)$. Thus, there are three hypermultiplets in the
two spinor representations ${\bf32}$ and ${\bf\o{32}}$ of $SO(12)$.
Nevertheless, the $14$ singlets may even form a further extension by a
nonabelian factor $G$. In principle, starting with a general CYM different
from $K_3\times T^2$ it can also happen that $(0,4)$ models occur.
Now, let us come to gauge groups of smaller rank.
\meno
$\bullet$ {\bf Gauge group $SU(6)$}
\smno
Starting with a model like in Table 4.1 the gauge group is at least
$SU(5)$. Analogously to the former case there occur extensions to larger
gauge groups, in particular to $SO(10)$ and $E_6$. However, it can also
happen that the gauge group is $SU(6)$. Including the simple current
$$ \Upsilon= \left(\Phi^3_{0,-1}\right)^2\, \left( \Phi^0_{0,0}\right)^3\,
  \left(\Phi^{U(1)_2}_{1,2}\right)^2\,\Phi^{SO(6)}_0 \eqno(7.9)$$
in (4.13) one obtains the massless $SU(5)$ spectrum shown in Table 7.5.
\bigno
\cl{\vbox{
\hbox{\vbox{\offinterlineskip
\def\tablespace{height2pt&\omit&&\omit&&\omit&&\omit&&\omit&&\omit&\cr}
\def\tablerule{\tablespace\noalign{\hrule}\tablespace}
\hrule\halign{&\vrule#&\strut\hskip0.2cm\hfil#\hfill\hskip0.2cm\cr
\tablespace
& && ${\bf0}$ && ${\bf10}$ && ${\bf\o{10}}$ && ${\bf5}$ && ${\bf\o{5}}$&\cr
\tablerule\tablerule
& spin $0$ && $348$ && $54$ && $4$ && $69$ && $119$ & \cr
\tablerule
& spin $1$ && $8$ && $0$ && $0$ && $1$ && $1$ & \cr
\tablespace}\hrule}}
\hbox{\hskip 0.5cm Table 7.5 \hskip 0.5cm $SU(6)$ model}}}
\meno
Thus, there are further gluons in the ${\bf5}$ and ${\bf\o5}$
representation of $SU(5)$ which on account of
$$ {\bf35}={\bf24}+{\bf5}+{\bf\o5}+{\bf1} \eqno(7.10)$$
leads to the enhanced gauge group $SU(6)$.
\meno
$\bullet$ {\bf Gauge group $SU(5)$}
\smno
A model with $SU(5)$ gauge group can be achieved by choosing the following
two simple currents:
$$\eqalignno{\Upsilon_1&=\Phi^3_{0,-1} \left( \Phi^0_{0,0}\right)^4\,
     \Phi^{U(1)_2}_{1,2} \Phi^{U(1)_2}_{0,2}\,\Phi^{SO(6)}_0 &(7.11)\cr
     \Upsilon_2&=\left(\Phi^3_{0,-1}\right)^2\,\left(\Phi^0_{0,0}\right)^3
     \,\left(\Phi^{U(1)_2}_{1,2}\right)^2\, \Phi^{SO(6)}_0.&\cr }$$
The massless spectrum is listed in Table 7.6.
\bigno
\cl{\vbox{
\hbox{\vbox{\offinterlineskip
\def\tablespace{height2pt&\omit&&\omit&&\omit&&\omit&&\omit&&\omit&\cr}
\def\tablerule{\tablespace\noalign{\hrule}\tablespace}
\hrule\halign{&\vrule#&\strut\hskip0.2cm\hfil#\hfill\hskip0.2cm\cr
\tablespace
& && ${\bf0}$ && ${\bf10}$ && ${\bf\o{10}}$ && ${\bf5}$ && ${\bf\o{5}}$&\cr
\tablerule\tablerule
& spin $0$ && $338$ && $64$ && $0$ && $55$ && $119$ & \cr
\tablerule
& spin $1$ && $10$ && $0$ && $0$ && $0$ && $0$ & \cr
\tablespace}\hrule}}
\hbox{\hskip 0.5cm Table 7.6 \hskip 0.5cm $SU(5)$ model}}}
\meno
The gauge anomaly cancellation condition for $SU(5)$ requires the same
number of chiral fermions in the ${\bf10}$ and ${\bf\o5}$ representation.
This condition is satisfied in our example yielding a net number of
$N_{gen}=64$ generations which is not divisible by $5$. However, there
exists a model with $N_{gen}=320$, so that we expect our model to be an
orbifold of that model. It is a general problem of our construction that
we do not have control over the geometric interpretation of dividing out
a certain set of simple currents. It can correspond either to an orbifold
construction or to a new vector bundle.
\pano
There is another example which shows that the number of generations in
${\bf\o{10}}$ does not need to be zero. Therefore, we consider the tensor
product of minimal models $(1)(4)^4$ also adding up to $c=9$. This model
has already turned out to yield a lot of different massless spectra
[\sche]. For instance, choosing the two simple currents
$$\eqalignno{\Upsilon_1&=\Phi^1_{0,-1}\left( \Phi^0_{0,0}\right)^4\,
     \Phi^{U(1)_2}_{-1,2} \Phi^{U(1)_2}_{0,2}\, \Phi^{SO(6)}_v &(7.12)\cr
     \Upsilon_2&=\Phi^0_{0,0} \Phi^4_{3,-1} \left( \Phi^0_{0,0}\right)^3\,
     \left(\Phi^{U(1)_2}_{1,2}\right)^2\, \Phi^{SO(6)}_0 &\cr }$$
one obtains the spectrum presented in Table 7.7.
\bigno
\cl{\vbox{
\hbox{\vbox{\offinterlineskip
\def\tablespace{height2pt&\omit&&\omit&&\omit&&\omit&&\omit&&\omit&\cr}
\def\tablerule{\tablespace\noalign{\hrule}\tablespace}
\hrule\halign{&\vrule#&\strut\hskip0.2cm\hfil#\hfill\hskip0.2cm\cr
\tablespace
& && ${\bf0}$ && ${\bf10}$ && ${\bf\o{10}}$ && ${\bf5}$ && ${\bf\o{5}}$&\cr
\tablerule\tablerule
& spin $0$ && $386$ && $51$ && $3$ && $66$ && $114$ & \cr
\tablerule
& spin $1$ && $10$ && $0$ && $0$ && $0$ && $0$ & \cr
\tablespace}\hrule}}
\hbox{\hskip 0.5cm Table 7.7 \hskip 0.5cm Nonzero generation number in
  $\bf\o{10}$}}}
\meno
A stochastic search for net numbers of generations of the
corresponding Distler/Kachru model $(1)(4)^4$ yields only numbers
divisible by 3. In particular, $48$ net generations occur.
It is clear, that the smaller the rank of the generic gauge group
the larger the set of possible extensions of the gauge group. For
$SU(3)\times SU(2)$ models we find also extensions to the
semisimple groups $SU(6)\times SU(2)$ and $SU(4)\times SU(2)$.
\meno
$\bullet$ {\bf Gauge group $SU(6)\times SU(2)$}
\smno
For the quintic $(3)^5$ the simple current
$$ \Upsilon=\left(\Phi^3_{0,-1}\right)^3\, \left( \Phi^0_{0,0}\right)^2\,
  \left(\Phi^{U(1)_2}_{1,2}\right)^3\,\Phi^{SO(4)}_0 \eqno(7.13)$$
yields the $E_3$ spectrum in Table 7.8.
\bigno
\cl{\vbox{
\hbox{\vbox{\offinterlineskip
\def\tablespace{height2pt&\omit&&\omit&&\omit&&\omit&&\omit&&\omit&&\omit
 &\cr}
\def\tablerule{\tablespace\noalign{\hrule}\tablespace}
\hrule\halign{&\vrule#&\strut\hskip0.2cm\hfil#\hfill\hskip0.2cm\cr
\tablespace
& && ${\bf0}$ && ${\bf2}$ && ${\bf3}$ && ${\bf\o{3}}$ && ${\bf6}$ &&
 ${\bf\o{6}}$ &\cr
\tablerule\tablerule
& spin $0$ && $442$ && $170$ && $81$ && $181$ && $50$ && $0$ & \cr
\tablerule
& spin $1$ && $13$ && $0$ && $3$ && $3$ && $0$ && $0$ &\cr
\tablespace}\hrule}}
\hbox{\hskip 0.5cm Table 7.8 \hskip 0.5cm $SU(6)\times SU(2)$
model}}}
\meno
As expected by gauge anomaly cancellation for the $SU(3)$ factor the
number $\left(\#({\bf3})-\#({\bf\o{3}})\right)$ is twice the number
$\left(\#({\bf\o 6})-\#({\bf6})\right)$.
\meno
$\bullet$ {\bf Gauge group $SU(4)\times SU(2)$}
\smno
Analogously to the $SU(5)$ case one can obtain smaller gauge groups by
including more simple currents. For instance,
$$ \eqalignno{\Upsilon_1&=\left(\Phi^3_{0,-1}\right)^2\,
  \left(\Phi^0_{0,0}\right)^3\,\left(\Phi^{U(1)_2}_{1,2}\right)^2
  \Phi^{U(1)_2}_{0,2}\,\Phi^{SO(4)}_0&(7.14)\cr
  \Upsilon_2&=\left(\Phi^3_{0,-1}\right)^3\,\left(\Phi^0_{0,0}\right)^2\,
  \left(\Phi^{U(1)_2}_{1,2}\right)^3\,\Phi^{SO(4)}_0 &\cr }$$
gives the spectrum in Table 7.9.
\bigno
\cl{\vbox{
\hbox{\vbox{\offinterlineskip
\def\tablespace{height2pt&\omit&&\omit&&\omit&&\omit&&\omit&&\omit&&\omit
 &\cr}
\def\tablerule{\tablespace\noalign{\hrule}\tablespace}
\hrule\halign{&\vrule#&\strut\hskip0.2cm\hfil#\hfill\hskip0.2cm\cr
\tablespace
& && ${\bf0}$ && ${\bf2}$ && ${\bf3}$ && ${\bf\o{3}}$ && ${\bf6}$ &&
 ${\bf\o{6}}$ &\cr
\tablerule\tablerule
& spin $0$ && $386$ && $140$ && $68$ && $148$ && $42$ && $2$ & \cr
\tablerule
& spin $1$ && $11$ && $0$ && $1$ && $1$ && $0$ && $0$ &\cr
\tablespace}\hrule}}
\hbox{\hskip 0.5cm Table 7.9 \hskip 0.5cm $SU(4)\times SU(2)$
model}}}
\meno
In order to reduce the gauge group to $E_3$ one has to include
three simple currents.
\meno
$\bullet$ {\bf Gauge group $SU(3)\times SU(2)\times G$}
\smno
Choosing
$$ \eqalignno{\Upsilon_1&=\Phi^3_{0,-1}\left(\Phi^0_{0,0}\right)^4\,
   \Phi^{U(1)_2}_{1,2}\left(\Phi^{U(1)_2}_{0,2}\right)^2\,\Phi^{SO(4)}_0
   &\cr
   \Upsilon_2&=\left(\Phi^3_{0,-1}\right)^2\,\left(\Phi^0_{0,0}\right)^3\,
   \left(\Phi^{U(1)_2}_{1,2}\right)^2\Phi^{U(1)_2}_{0,2}\,\Phi^{SO(4)}_0
   &(7.15)\cr
   \Upsilon_2&=\left(\Phi^3_{0,-1}\right)^3\,\left(\Phi^0_{0,0}\right)^2\,
   \left(\Phi^{U(1)_2}_{1,2}\right)^3\,\Phi^{SO(4)}_0&\cr }$$
gives the model with $50$ net generations listed in Table 7.10.
\bigno
\cl{\vbox{
\hbox{\vbox{\offinterlineskip
\def\tablespace{height2pt&\omit&&\omit&&\omit&&\omit&&\omit&&\omit&&\omit
 &\cr}
\def\tablerule{\tablespace\noalign{\hrule}\tablespace}
\hrule\halign{&\vrule#&\strut\hskip0.2cm\hfil#\hfill\hskip0.2cm\cr
\tablespace
& && ${\bf0}$ && ${\bf2}$ && ${\bf3}$ && ${\bf\o{3}}$ && ${\bf6}$ &&
 ${\bf\o{6}}$ &\cr
\tablerule\tablerule
& spin $0$ && $370$ && $134$ && $54$ && $154$ && $50$ && $0$ & \cr
\tablerule
& spin $1$ && $13$ && $0$ && $0$ && $0$ && $0$ && $0$ &\cr
\tablespace}\hrule}}
\hbox{\hskip 0.5cm Table 7.10 \hskip 0.5cm $SU(3)\times SU(2)$
model}}}
\meno
Let us compute the additional factor $G$ for phenomenological reasons.
Since $6$ of the $13$ singlet vector fields originate from the simple
currents $\Upsilon_i$, they do not commute with the remaining $7$ $U(1)$
factors. A detailed analysis shows that the gauge group contains a further
$SU(3)$ factor. Thus, the entire gauge group of this model is
$SU(3)\times SU(2)\times\left( U(1)\right)^{5}\times SU(3)$.
Besides the leptonic partners of the $50$ quarks there occur $84$
further Higgses in the $(0,2)$ representation of $E_3$.
\pano
What can also happen is the occurrence of gauge groups of rank higher
than six.
\meno
$\bullet$ {\bf Gauge group $E_7$}
\smno
For this extended group we again start with the model $(1)(4)^4$ and select
only the simple current
$$\Upsilon_1=\Phi^1_{0,-1}\left( \Phi^0_{0,0}\right)^4\,
     \Phi^{U(1)_2}_{1,2}\, \Phi^{SO(8)}_s. \eqno(7.16) $$
Due to the decomposition of the adjoint representation of $E_7$ into
irreducible representations of $SO(10)$
$$ {\bf133}={\bf45}+{\bf10}+{\bf10}+{\bf16}+{\bf16}+{\bf\o{16}}+
 {\bf\o{16}}+{\bf1}+{\bf1}+{\bf1}+{\bf1}\eqno(7.17)$$
the massless spectrum in Table 7.11 yields $36$ generations in the
${\bf56}$ representation of $E_7$.
\bigno
\cl{\vbox{
\hbox{\vbox{\offinterlineskip
\def\tablespace{height2pt&\omit&&\omit&&\omit&&\omit&&\omit&\cr}
\def\tablerule{\tablespace\noalign{\hrule}\tablespace}
\hrule\halign{&\vrule#&\strut\hskip0.2cm\hfil#\hfill\hskip0.2cm\cr
\tablespace
& && ${\bf0}$ && ${\bf10}$ && ${\bf16}$ && ${\bf\o{16}}$ &\cr
\tablerule\tablerule
& spin $0$ && $416$ && $72$ && $36$ && $36$ & \cr \tablerule
& spin $1$ && $7$ && $2$ && $2$ && $2$ & \cr
\tablespace}\hrule}}
\hbox{\hskip 0.5cm Table 7.11 \hskip 0.5cm $E_7$ model}}}
\meno
$N_{gen}=36$ can also be reproduced using (7.4).
Furthermore, one can even produce the gauge group $SU(7)$.
\meno
$\bullet$ {\bf Gauge group $SU(7)$}
\smno
Again we choose $(1)(4)^4$ and include the following two simple currents:
$$\eqalignno{\Upsilon_1&=\Phi^1_{0,-1}\left( \Phi^0_{0,0}\right)^4\,
     \Phi^{U(1)_2}_{1,2} \Phi^{U(1)_2}_{0,2}\, \Phi^{SO(6)}_s &(7.18)\cr
     \Upsilon_2&=\Phi^0_{0,0} \Phi^4_{3,-1} \left( \Phi^0_{0,0}\right)^3\,
     \left(\Phi^{U(1)_2}_{1,2}\right)^2\, \Phi^{SO(6)}_0 . &\cr }$$
The extension of $SU(5)$ to $SU(7)$ can be seen from Table 7.12 taking into
account the decomposition
$$    {\bf49}={\bf24}+{\bf5}+{\bf5}+{\bf\o 5}+{\bf\o 5}+{\bf1}+{\bf1}
     +{\bf1}+{\bf1} \eqno(7.19)$$
\bigno
\cl{\vbox{
\hbox{\vbox{\offinterlineskip
\def\tablespace{height2pt&\omit&&\omit&&\omit&&\omit&&\omit&&\omit&\cr}
\def\tablerule{\tablespace\noalign{\hrule}\tablespace}
\hrule\halign{&\vrule#&\strut\hskip0.2cm\hfil#\hfill\hskip0.2cm\cr
\tablespace
& && ${\bf0}$ && ${\bf10}$ && ${\bf\o{10}}$ && ${\bf5}$ && ${\bf\o{5}}$&\cr
\tablerule\tablerule
& spin $0$ && $463$ && $49$ && $7$ && $90$ && $132$ & \cr
\tablerule
& spin $1$ && $15$ && $0$ && $0$ && $2$ && $2$ & \cr
\tablespace}\hrule}}
\hbox{\hskip 0.5cm Table 7.12 \hskip 0.5cm $SU(7)$ model}}}
\meno
A net number of $42$ generations also occur for the Distler/Kachru models.
Finally, we present an example with four net generations and gauge
group $SO(10)$. This spectrum results from the quintic $(3)^5$ and the
simple currents
$$\eqalignno{\Upsilon_1&=\Phi^3_{-2,-1}\left( \Phi^3_{-1,2}\right)^2\,
     \Phi^3_{-3,0}\Phi^0_{0,0}\,
     \Phi^{U(1)_2}_{1,2}\, \Phi^{SO(8)}_c &(7.20)\cr
     \Upsilon_2&=\Phi^0_{0,0} \Phi^3_{-3,0} \left( \Phi^0_{0,0}\right)^2\,
     \Phi^0_{1,1}\,\Phi^{U(1)_2}_{1,2}\,\Phi^{SO(8)}_v &\cr }$$
with the spectrum listed in Table 7.13. This model is likely an orbifold of
a Distler/Kachru model.
\bigno
\cl{\vbox{
\hbox{\vbox{\offinterlineskip
\def\tablespace{height2pt&\omit&&\omit&&\omit&&\omit&&\omit&\cr}
\def\tablerule{\tablespace\noalign{\hrule}\tablespace}
\hrule\halign{&\vrule#&\strut\hskip0.2cm\hfil#\hfill\hskip0.2cm\cr
\tablespace
& && ${\bf0}$ && ${\bf10}$ && ${\bf16}$ && ${\bf\o{16}}$ &\cr
\tablerule\tablerule
& spin $0$ && $254$ && $32$ && $18$ && $14$ & \cr \tablerule
& spin $1$ && $7$ && $0$ && $0$ && $0$ & \cr
\tablespace}\hrule}}
\hbox{\hskip 0.5cm Table 7.13 \hskip 0.5cm Four net generations}}}
\meno
All these examples show that one can really find suitable simple currents
$\Upsilon$ breaking both the left moving $N=2$ supersymmetry and the gauge
group $E_6$. Furthermore, the massless spectra obtained can all be found
in the context of $(0,2)$ (non)linear $\sigma$ models encouraging further
investigations about their relationship.
\vfill
\eject
\bigno
\section{8.\ Conclusion}
\meno
In this paper we have presented a method to construct modular invariant
partition functions of four space-time dimensional heterotic string vacua
with $(0,2)$ world sheet supersymmetry and generic gauge groups $E_r$ with
$3\leq r\leq6$. This constructively proves the existence of bona fide CFTs
with all the properties known for the conformal fixed points of $(0,2)$
supersymmetric CYM/LG models. In particular, these vacua are not suffering
from destabilizing instanton corrections. Clearly, our construction is not
unique and there might exist different construction schemes of $(0,2)$
CFTs, especially those starting with a truly heterotic modular invariant
partition function. However, $(2,2)$ models have taught us that due to
the GSO projection the spectra obtained are highly degenerate. Thus, we
hope that more than only a very small subset of all exactly solvable
$(0,2)$ vacua can be realized by the simple current method.
\pano
Since both orbifolds of the CYM and the choice of a more general stable
vector bundle for the left moving $\sigma$ model fermions are encoded in
the same manner in this class of CFTs, a direct correspondence between
simple currents and the defining data of the latter bundles might be
hardly to reveal. One ansatz to construct such a map could be the
coincidence of the model in Table 7.1 and a complete intersection in a
weighted projective space. If such a one to one identification
could be achieved for at least a few models, it would be possible to
calculate and compare further properties of physical importance. For
instance, one could address questions like the exactness of a first order
calculation of some of the Yukawa couplings in the CYM/LG framework.
Furthermore, since the class of exactly solvable models exhibits exact
mirror symmetry, one expects such an isomorphism for the CYM/LG
formulation, as well.
\pano
Thus, in order to learn more about the degeneracy of the spectra and the
appearing net numbers of generations one has to extend further the set of
explicitly known models [\selfb].
\meno
{\bf Acknowledgements}
\smno
It is a pleasure to thank L.\ Dolan and W.\ Nahm for discussion, S.\ Kachru
for his hint on the coincidence concerning the model in Table 7.1,
S.\ Chaudhuri for interesting questions on extensions of the gauge groups
and in particular R.\ Schimmrigk for pointing our attention to string
models with $(0,2)$ world sheet supersymmetry. This work is supported by
U.S.\ DOE grant No.\ DE-FG05-85ER-40219.
\bigno
\section{References}
\meno
\bibitem{\free}
I.\ Antoniadis, C.\ Bachas and C.\ Kounnas,
 Nucl.\ Phys.\ {\bf B289} (1987) 87,
L.\ Ib\'a$\tilde{\rm n}$ez, H.\ Nilles and F.\ Quevedo,
 Phys.\ Lett.\ {\bf B187} (1987) 25,
H.\ Kawai, D.\ Lewellen and S.\ Tye,
 Nucl.\ Phys.\ {\bf B288} (1987) 1,
W.\ Lerche, D.\ L\"ust and A.N.\ Schellekens,
 Nucl.\ Phys.\ {\bf B287} (1987) 477,
K.\ Narain, M.\ Sarmadi and C.\ Vafa,
 Nucl.\ Phys.\ {\bf B288} (1987) 551
\bibitem{\banks} T.\ Banks, L.J.\ Dixon, D.\ Friedan and E.\ Martinec,
{\it Phenomenology and conformal field theory, or Can string theory predict
the weak mixing angle}?,
Nucl.\ Phys.\ {\bf B299} (1988) 613
\bibitem{\self} R.\ Blumenhagen and A.\ Wi{\ss}kirchen, {\it Generalized
string functions of $N=1$ space-time supersymmetric string vacua},
Phys.\ Lett.\ {\bf B349} (1995) 63
\bibitem{\selfb} R.\ Blumenhagen and A.\ Wi{\ss}kirchen,
{\it work in progress}
\bibitem{\cande} P.\ Candelas, G.T.\ Horowitz, A.\ Strominger and
E.\ Witten,
{\it Vacuum configurations for superstrings},
Nucl.\ Phys.\ {\bf B258} (1985) 46
\bibitem{\dine} M.\ Dine, N.\ Seiberg, X.G.\ Wen and E.\ Witten,
{\it Nonperturbative effects on the string world sheet I+II},
Nucl.\ Phys.\ {\bf B278} (1986) 769, Nucl.\ Phys.\ {\bf B289} (1987) 319
\bibitem{\disgre} J.\ Distler and B.\ Greene,
{\it Aspects of $\,(2,0)$ string compactifications},
Nucl.\ Phys.\ {\bf B304} (1988) 1
\bibitem{\diskacha} J.\ Distler and S.\ Kachru,
{\it $(0,2)$ Landau-Ginzburg theory},
Nucl.\ Phys.\ {\bf B413} (1994) 213
\bibitem{\diskachb} J.\ Distler and S.\ Kachru,
{\it Singlet couplings and $(0,2)$ models},
Nucl.\ Phys.\ {\bf B430} (1994) 13
\bibitem{\egu} T.\ Eguchi, H.\ Ooguri, A.\ Taormina and S.K.\ Yang,
{\it Superconformal algebras and string compactification on manifolds
with $SU(n)$ holonomy},
Nucl.\ Phys.\ {\bf B315} (1989) 193
\bibitem{\gepe} D.\ Gepner, {\it Space-time supersymmetry in
compactified string theory and superconformal models},
Nucl.\ Phys.\ {\bf B296} (1988) 757
\bibitem{\kava} S.\ Kachru and C.\ Vafa,
{\it Exact results for $N=2$ compactifications of heterotic strings},
preprint HUTP-95/A016 (hep-th/9505105)
\bibitem{\oda} S.\ Odake, {\it Character formulas of an
extended superconformal algebra relevant to string compactification},
Int.\ J.\ Mod.\ Phys.\ {\bf A5} (1990) 897
\bibitem{\schwar} A.N.\ Schellekens and N.P.\ Warner,
{\it Anomalies, characters and strings},
Nucl.\ Phys.\ {\bf B287} (1987) 317
\bibitem{\sche} A.N.\ Schellekens and S.\ Yankielowicz,
{\it Extended chiral algebras and modular invariant partition functions},
Nucl.\ Phys.\ {\bf B327} (1989) 673
\bibitem{\schz} A.N.\ Schellekens and S.\ Yankielowicz,
{\it Modular invariants from simple currents. An explicit proof},
Phys.\ Lett.\ {\bf B227} (1989) 387
\bibitem{\schd} A.N.\ Schellekens and S.\ Yankielowicz,
{\it New modular invariants for $N=2$ tensor products and four-dimensional
strings}, Nucl.\ Phys.\ {\bf B330} (1990) 103
\bibitem{\silver} E.\ Silverstein and E.\ Witten,
{\it Criteria for conformal invariance of $\,(0,2)$ models},
Nucl.\ Phys.\ {\bf B444} (1995) 161
\bibitem{\witten} E.\ Witten,
{\it New issues in manifolds of $SU(3)$ holonomy},
Nucl.\ Phys.\ {\bf B268} (1986) 79
\bibitem{\wit} E.\ Witten,
{\it Phases of $N=2$ theories in two dimensions},
Nucl.\ Phys.\ {\bf B403} (1993) 159
\vfill
\end